\documentclass[pre,twocolumn,superscriptaddress,showpacs]{revtex4}
\usepackage{epsfig}
\usepackage{graphicx}
\usepackage{amssymb, amstext, amsmath}
\usepackage{hyperref}
\usepackage{verbatim}
\usepackage{color}
\usepackage{bbold}
\usepackage[normalem]{ulem}

\begin{document}
\title{Non-Canonical Statistics of a Spin-Boson Model:
Theory and Exact Monte-Carlo Simulations}
\author{Chee Kong Lee} \email{cqtlck@nus.edu.sg}
\affiliation{Centre for
Quantum Technologies, National University of Singapore, 117543, Singapore}
\author{Jianshu Cao} \email{jianshu@mit.edu}
\affiliation{Department of Chemistry, Massachusetts Institute of Technology, Cambridge, Massachusetts 02139, USA}
\author{Jiangbin Gong}\email{phygj@nus.edu.sg}
\affiliation{Department of Physics and Center for Computational
Science and Engineering, National University of Singapore, 117542, Singapore}
\affiliation{NUS Graduate School for Integrative Sciences
and Engineering, Singapore 117597, Singapore}
\begin{abstract}
Equilibrium canonical distribution in statistical mechanics assumes weak system-bath coupling (SBC).
In real physical situations this assumption can be invalid and equilibrium quantum statistics of the system
may be non-canonical. By exploiting both polaron transformation and perturbation theory in a spin-boson model, an analytical treatment is advocated
to study non-canonical statistics of a two-level system at arbitrary temperature and for arbitrary SBC strength, yielding
theoretical results in agreement with exact Monte-Carlo simulations.  In particular,
the eigen-representation of system's reduced density matrix is used to quantify non-canonical statistics as well as the quantumness of the open system.
For example, it is found that irrespective of SBC strength, non-canonical statistics enhances as temperature decreases but vanishes at high temperature.
\end{abstract}



\pacs{03.65.Yz; 03.67.Mn; 05.30.-d} \date{\today} \maketitle
\section{Introduction}
Equilibrium canonical distribution is a fundamental result in statistical mechanics, but with
an implicit assumption that the system-bath coupling (SBC) is vanishingly weak.
In real physical situations, such as light-harvesting systems ~\cite{Brixner2005}, super-conducting qubits \cite{You2011}, and atom-cavity systems ~\cite{Auer2011},
this weak SBC assumption can be invalid.  Then a good separation between the system and the bath is lost, leading to
non-canonical equilibrium statistics for the system (though the system plus the bath as a whole still satisfies a canonical distribution).
At present non-canonical statistics is a subject of great interest.  It challenges
intuitive concepts in statistical mechanics \cite{heatcapacity-note1} and brings about new understandings of open quantum systems.
Because canonical statistics can be recovered in
the classical limit for a wide class of microscopic open-system models \cite{mean-force-note2},
non-canonical statistics can also serve as an indicator of the quantumness of open systems.

To better understand non-canonical statistics and for future bench-marking purposes, an analytical treatment of non-canonical statistics at arbitrary
temperature and also for the whole range of SBC strength is important.  This task is achieved here in a spin-boson model by exploiting both the polaron transformation technique and perturbation theory, with results applicable to
almost the entire parameter space (certain conditions to be elaborated below).   Though our theory
does provide all the information of system's reduced density matrix (RDM),  one simple measure is employed
to quantify the degree of non-canonical statistics.  Specifically, we use the angle by which the diagonal representation of RDM is rotated away from
the energy eigen-representation of the system as the measure. Evidently, this angle is zero for a canonical distribution because the associated RDM is diagonal in
energy basis states.  It is then shown how an increasing SBC strength or a decreasing temperature strengthens non-canonical
statistics.  Interestingly, at sufficiently high temperature, our theory reveals that the standard canonical statistics will be recovered asymptotically.
The theoretical results are found to be in full agreement with exact Monte-Carlo simulations. The observation that non-canonical statistics vanishes in the high temperature regime
resonates with the view that non-canonical statistics
reflects the quantumness of the open system, which is expected to diminish at high temperature.
Furthermore, we exploit our simple non-canonical statistics measure to identify an interesting temperature scale,
at which the non-canonical equilibrium statistics is most sensitive to temperature variations.

\section{Theory}
The total Hamiltonian $H_{t}$ of a system plus a bath is written as
\begin{eqnarray}
	H_t=H_S+H_B+ H_{SB},
\end{eqnarray}
where the three terms represent Hamiltonians of the system, the bath, and SBC, respectively.  The system and the bath as a whole is assumed to be in contact with a super-reservoir
at temperature $T$.
Long after quantum relaxation takes place, the system and the bath as a whole reaches a canonical equilibrium state $\rho_t$.
 The equilibrium RDM of the system
is given by~\cite{weiss2008}
\begin{eqnarray} 
		\rho_S= \mbox{tr}_B[\rho_t] =\frac{\mbox{tr}_B[\mbox{e}^{-\beta H_t}]}{\mbox{tr}[\mbox{e}^{-\beta H_t}]}, \label{eqn:RDM}
\end{eqnarray}
where $\beta=\frac{1}{k_B T}$, $\mbox{tr}[\cdot]$ denotes the trace over the system as well as the bath, and $\mbox{tr}_B[\cdot]$ denotes the
trace over the bath only.  Due to the non-vanishing contribution of $H_{SB}$ to $H_{t}$, in general the equilibrium RDM
is no longer a canonical distribution of the system alone~\cite{Grabert1988}. It is hence of importance to develop a theoretical treatment to
 systematically investigate the dependence of
$\rho_S$ upon temperature $T$ and the magnitude of $H_{SB}$.
We achieve this task here for a spin-boson model with the Hamiltonian (we set $\hbar =1$)
\begin{eqnarray} 
	H_t=\frac{\epsilon}{2} \sigma_z + \frac{\Delta}{2} \sigma_x + \sum_k
   \omega_k b^\dagger_k b_k + \sigma_z \sum_k g_k(b^\dagger_k + b_k),
\end{eqnarray}
where $\sigma_i$ ($i=x, y, z$) are the usual Pauli matrices, $\epsilon$ is the
energy splitting between two levels of the system, and $\Delta$ is the tunneling matrix
element. The bath is modeled as a set of harmonic oscillators with frequencies $\omega_k$,
and the couplings to the spin system are denoted by $g_k$. The properties of the bath are fully characterized by its
spectral density, namely, $J(\omega)=\pi \sum_k g_{k}^2 \delta(\omega-\omega_k)$.   For a fully analytical development
we first assume a super-ohmic spectral density with exponential cut-off, i.e.,
$J(\omega) =\gamma \omega^3 \mbox{e}^{-\omega/\omega_c}$, where $\gamma$
characterizes the SBC strength and $\omega_c$ is the cut-off frequency of the bath. Though such an open-system model
is standard and (deceptively) simple, even its stationary properties
do not have exact solutions (not to mention its dynamics). Indeed, only the case of a single-mode spin-boson model with $\epsilon=0$ can be regarded as integrable~\cite{Braak2011}.
As such, perturbation theory becomes one of the few options. Note, however, a naive perturbation theory would fail
as our central concern is beyond the weak SBC regime.

To capture the effects of a finite $H_{SB}$, we exploit a standard paloron transformation as our starting step. To that end,
we transform the total Hamiltonian $H_t$ to $\tilde{H}_t= \mbox{e}^{F} H_t \mbox{e}^{-F}$ (tildes denote operators in the transformed
polaron picture) where
\begin{eqnarray}
F\equiv \sigma_z \sum_k \frac{g_k}{\omega_k}( b^\dagger_k -b_k).
\end{eqnarray}
  We then obtain (up to a constant)%
\begin{eqnarray} 
	 \tilde{H}_t &=& \tilde{H}_S +\tilde{H}_B + \tilde{H}_{SB}, \nonumber \\
	 &=&\frac{\epsilon}{2} \sigma_z + \frac{\Delta_R}{2} \sigma_x  + \sum_k \omega_k b_k^\dagger b_k    +    	\sigma_x V_x + \sigma_y V_y,
\end{eqnarray}
with  $\tilde{H}_S\equiv  \frac{\epsilon}{2} \sigma_z + \frac{\Delta_R}{2} \sigma_x$,  $\tilde{H}_B=H_B$,
and $\tilde{H}_{SB}=\sigma_x V_x + \sigma_y V_y$. Two remarks are in order: (i) for $\tilde{H}_S$,
the $\sigma_x$ (tunneling) term is now renormalized by SBC, because
\begin{eqnarray}
\Delta_R/\Delta = R =
 \exp\Big[-2 \int^\infty_0 \frac{d\omega}{\pi}\frac {J(\omega)}{\omega^2} \coth(\beta \omega/2)\Big].
 \end{eqnarray}
  Using
 the super-ohmic $J(\omega)$ mentioned above, we obtain $R=\exp[- \frac{2 \gamma}{\pi \beta^2}(2 \psi'(1/\beta \omega_c) - \omega_c^2 \beta^2) ]$,
 with $\psi'$
 being the derivative of the digamma function;
and (ii)  $\tilde{H}_{SB}$, the SBC in the paloron picture, assumes
a form very different from $H_{SB}$, because the bath operators entering into SBC are now
\begin{eqnarray}
V_x&=&\frac{\Delta}{4} (D_{+}^2 + D_{-}^2 - 2R), \\
V_y& = &\frac{\Delta}{4i} (D_{-}^2 -D_{+}^2),
\end{eqnarray}
 with
 $D_{\pm}=\exp \big[\pm\frac{g_k}{\omega_k}(b_k^\dagger -b_k)\big]$ and $\langle D_{\pm}^2 \rangle_{H_B}=R$~\cite{correlation-function-note}. We stress that so far our procedure is exact.
As seen below, this polaron transformation is advantageous in treating cases of non-vanishing SBC strength
because the correlation functions of $V_x$ and of $V_y$  decay faster for a larger $\gamma$.
For later use we also introduce density matrices in the polaron picture, i.e.,
$\tilde{\rho}=\mbox{e}^{F} \rho \mbox{e}^{-F}$, where $\rho$ can be the RDM
$\rho_S$ or the full density matrix $\rho_t$.  For example,
$\tilde{\rho}_{t}=\frac{\mbox{e}^{-\beta \tilde{H}_t}}{\mbox{tr}[\mbox{e}^{-\beta \tilde{H}_t}]}$, and
$\tilde{\rho}_{S}=\mbox{tr}_B[\tilde{\rho}_t]$.

We now proceed with the calculations of the matrix elements of $\rho_S$, expressed in terms of the $\sigma_z$ basis.
The diagonal elements $\rho_S^{11}$ and $\rho_S^{22}$ are given by
$\frac{1}{2}(1\pm \mbox{tr}_S[\sigma_z\rho_S])$ and  the off-diagonal element $\rho_S^{12}$ is given by $\mbox{tr}_S[\sigma_{-}\rho_S]=
\mbox{tr}[\sigma_{-}\rho_t]$.
Noticing that $\mbox{tr}_S [\sigma_z \rho_S]= \mbox{tr}_S [\sigma_z \tilde{\rho}_S]$, one may obtain the diagonal elements  directly if
$\tilde{\rho}_S$, the RDM in the polaron picture, is analytically known. On the other hand, because $\sigma_{-}$
operator does not commute with the polaron transformation operator $F$ defined above, calculating $\rho_S^{12}$ in the polaron picture
is more involving. Nevertheless, we find
\begin{eqnarray}
\rho_{S}^{12}=\mbox{tr}[\sigma_{-}\rho_t]=\mbox{tr}[\tilde{\sigma}_{-}\tilde{\rho}_t]=  \mbox{tr}[{\sigma}_{-}D_{-}^2\tilde{\rho}_t],
\end{eqnarray}
indicating that $\rho_{S}^{12}$ can be still obtained from $\tilde{\rho}_t$, but not from $\tilde{\rho}_S$.

Having observed how all RDM elements may be evaluated in the polaron picture, we turn to the central expression $\mbox{e}^{-\beta\tilde{H}_t}$
in $\tilde{\rho}_t$.  We treat $\beta$ as an imaginary time and exploit a ``time"-dependent perturbation theory in terms of $\tilde{H}_{SB}$.
We first briefly mention the calculations of the RDM diagonal elements, a subject of a recent technical study by us~\cite{Lee2012}.
In particular, due to the fact that $\mbox{tr}_B[\tilde{H}_{SB}]=0$, the leading-order contribution of $\tilde{H}_{SB}$
is of the second order, from which we have $\tilde{\rho}_S \approx \tilde{\rho}_{S,(0)} + \tilde{\rho}_{S,(2)}$, where~\cite{Geva2000}
 \begin{eqnarray} 
	\tilde{\rho}_{S,(0)} &= & \frac{\mbox{e}^{-\beta \tilde{H}_S}}{ Z_{S,(0)}};\label{zerothorder}   \\
    \tilde{\rho}_{S,(2)} &= & \frac{A}{Z_{S,(0)}} - \frac{Z_{S,(2)}}{[Z_{S,(0)}]^2} \mbox{e}^{-\beta \tilde{H}_S},
	\label{second-order-state}
\end{eqnarray} with
\begin{eqnarray}
A=\sum_{n=x,y} \int^\beta _0 d\tau  \int^{\tau}_0 d\tau'  C_{n}(\tau-\tau') \mbox{e}^{-\beta\tilde{H}_S} \sigma_n(\tau) \sigma_n(\tau'),
\end{eqnarray}
 $Z_{S,(0)}= \mbox{tr}_S[\mbox{e}^{-\beta \tilde{H}_S}]$ and $Z_{S,(2)}=\mbox{tr}_S[A]$.  In addition,
the operators in imaginary time are defined as
$O(\tau)\equiv \mbox{e}^{\tau \tilde{H}_0} O \mbox{e}^{-\tau \tilde{H}_0}$, where
$\tilde{H}_0= \tilde{H}_S + \tilde{H}_B$.
The bath correlation functions $C_{n}(\tau)=\langle V_n(\tau) V_n \rangle_{H_B}$ are given by
\begin{eqnarray}
C_x(\tau) &= &\frac{\Delta_R^2}{8}(\mbox{e}^{\phi(\tau)} + \mbox{e}^{-\phi(\tau)}-2), \\
C_y(\tau)&= &\frac{\Delta_R^2}{8}(\mbox{e}^{\phi(\tau)} - \mbox{e}^{-\phi(\tau)}),
\end{eqnarray}
 with $\phi(\tau) = 4  \int^\infty_0 \frac{d\omega}{\pi} \frac{J(\omega)}{\omega^2} \frac{\cosh[\frac{1}{2}(\beta -2\tau) \omega]}{\sinh(\beta  \omega/2)}$~\cite{phi-note3}.
A straightforward inspection of these expressions (in particular, the term $\Delta_R^2 \mbox{e}^{\phi(\tau)})$
shows that as $\gamma$ (the SBC strength) increases, the
correlation functions $C_n(\tau)$ always decreases exponentially with $\gamma$,
thus enhancing the perturbation theory in imaginary time. Indeed, in the strong SBC coupling limit, the second-order
correction  $\tilde{\rho}_S^{(2)}$ approaches zero.  This roughly explains how our polaron-picture-based perturbation theory, by construction, may be suitable for
treating finite SBC strength. It is now also clear that our theory is perturbative in terms of $\tilde{H}_{SB}$, but non-perturbative
in terms of $H_{SB}$.  We also note that though the integral expression in Eq.~(\ref{second-order-state}) is complicated, it involves $2\times 2$
matrices only. The final expression for the trace $\mbox{tr}_S[\sigma_z \tilde{\rho}_S]$ is hence also analytical.
Detailed calculations~\cite{Lee2012} show that the RDM diagonal elements thus obtained are accurate so long as the tunneling element $\Delta$ is not large as compared to the bath cut-off frequency $\omega_c$ (therefore not a slow bath).

The crucial task in theory here is to explicitly evaluate the off-diagonal element of $\rho_{S}$ via $\mbox{tr}[{\sigma}_{-}D_{-}^2\tilde{\rho}_t]$.
In this case, due to the correlation between the system and the bath, the first-order contribution of $\tilde{H}_{SB}$ to $\mbox{e}^{-\beta \tilde{H}_t}$ and
hence to $\tilde{\rho}_t$ are already nonzero (upon thermal averaging).  As such, it suffices to consider a first-order perturbation theory in imaginary time
for the total density in the polaron picture. With some details elaborated in Appendices \cite{supple},
we finally obtain $\rho^{12}_{S} \approx \rho^{12}_{S,(0)} + \rho^{12}_{S,(1)}$, with \begin{eqnarray} 
		\rho^{12}_{S,(0)} &=& - \frac{R \Delta_R }{2\eta} \tanh(\beta \eta /2); \label{o1} \\
	  \rho^{12}_{S,(1)} &=& -\sum_{n=x,y} \int^\beta_0 d\tau S_n(\tau) K_n(\tau),      \label{o2}
\end{eqnarray}
$\eta=\sqrt{\epsilon^2 + \Delta_R^2}$. Here
 $S_n(\tau)=\langle \sigma_n(\tau) \sigma_- \rangle_{\tilde{H}_S}$ and
 $K_n(\tau)=\langle V_n(\tau) D_-^2 \rangle_{H_B}$ are the correlation functions
 of the system and of the bath, respectively. They are given by
 \begin{eqnarray}
S_x(\tau)&=&\frac{\Delta_R^2}{2 \eta^2} + \frac{\epsilon \mbox{sech}(\beta \eta /2)}{2 \eta^2}\nonumber \\
&&\times \Big(\epsilon \cosh[\frac{1}{2}(\beta - 2 \tau)\eta] + \eta \sinh[\frac{1}{2}(\beta - 2 \tau)\eta]\Big), \nonumber
\\ \ \\
S_y(\tau)&=&-\frac{i}{2}\mbox{sech}(\beta \eta /2) \nonumber \\
&&\times \Big( \cosh[\frac{1}{2}(\beta - 2 \tau)\eta] + \frac{\epsilon}{\eta} \sinh[\frac{1}{2}(\beta - 2 \tau)\eta]\Big). \nonumber \\
\end{eqnarray}
The bath correlation functions are
$K_x(\tau)= 2C_x(\tau)/\Delta$ and $K_y(\tau)= 2i C_y(\tau)/\Delta$.
Note that the first-order correction here is again linked with the above-defined bath correlation function $C_n(\tau)$.  So
by construction, our perturbation theory for off-diagonal elements of RDM works even better for stronger SBC coupling (thus analogous to the previous treatment
for  $\rho_S^{11}$ and $\rho_S^{22}$).

\section{Results}
\begin{figure}[h]  
	\center
	 \includegraphics[width=3.5in]{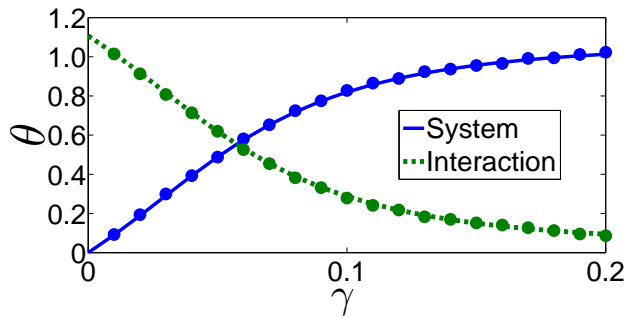}
         \caption{Coupling-strength dependence of the angle to be rotated on the Bloch sphere to reach eigenstates of $H_S$ (solid line) or $H_{SB}$ (dashed line) from eigenstates of
         equilibrium RDM, for $\beta=1$, $\epsilon=0.5$ and $\omega_c=5$ (in unit of $\Delta$). Solid dots are numerically exact Monte
         Carlo simulations results (for details of this method, see Ref.~\cite{Moix2012}).}
     \label{FIG:Angle1}
\end{figure}

With all the RDM elements analytically obtained above, we now
examine the validity of our theoretical results and reveal interesting physics in non-canonical equilibrium statistics.
Instead of examining all the RDM elements (one exception later), we use a single quantity to characterize non-canonical statistics, i.e.,
the smallest possible angle ($\theta$) to be rotated (in radians) on the Bloch sphere to reach the eigenstates of $H_S$ from the diagonal representation of the RDM.
The theoretical results (solid line) are plotted in Fig.~\ref{FIG:Angle1} as a function of the SBC strength, $\gamma$, for a fixed temperature.
For small values of $\gamma$, $\theta$ is small, so the RDM's diagonal representation is close to, or significantly overlapping with, that of $H_S$.
This is expected because for weak SBC strength, the equilibrium statistics should be canonical.
As $\gamma$ increases, $\theta$ increases, indicating that the RDM diagonal representation continuously and monotonously rotates away from
the eigenstates of $H_S$ \cite{footnote}. To further elucidate the continuous change in $\theta$, an analogous angle,
namely, the angle the RDM diagonal representation
should be rotated to reach the eigenstates of $H_{SB}$ is also plotted in Fig.~\ref{FIG:Angle1} (dashed line).
Interestingly, for large values of $\gamma$,  the RDM diagonal representation is seen to approach that of $H_{SB}$.
Indeed,   because $C_n(\tau)\rightarrow 0$ for large $\gamma$, all the perturbative corrections to $\rho^{11}_{S}$, $\rho^{22}_{S}$, and $\rho^{12}_S$ approach zero,
and hence the RDM approaches $\exp(-\frac{\epsilon}{2}\beta \sigma_z)$, whose diagonal representation should be parallel to that of $H_{SB}$ (as both
are a function of $\sigma_z$).  This is the case at arbitrary temperature.
Our theoretical results are also in quantitative agreement
with the solid dots shown in Fig.~\ref{FIG:Angle1}, obtained numerically from Monte Carlo simulations based on imaginary time path
integral (a powerful method if the bath temperature is not too low \cite{Moix2012}).  That is, for a varying SBC strength, either weak or strong,
our theory and numerically exact results agree.  This confirms that
our analytical treatment for RDM off-diagonal elements performs equally well in the
regime valid for treating the RDM diagonal elements (hence almost the entire parameter space ~\cite{Lee2012}).

In a different context, i.e., decoherence dynamics \cite{Zurek1981, Zurek2003},
the RDM diagonal representation is regarded as a special representation, often called a preferred basis of decoherence.  It is in this
special representation that decoherence can be understood as the disappearance of the off-diagonal matrix elements of a time-evolving
RDM.  A recent study using a low-dimensional quantum chaos model as a quantum bath \cite{Wang2012} shows that
the preferred basis of decoherence shows exactly the same qualitative
behavior as observed in Fig.~\ref{FIG:Angle1} for the equilibrium RDM, i.e., after a short period of decoherence
the preferred basis of a system
coincides with the eigenstates of $H_S$ for weak SBC and becomes the eigenstates of $H_{SB}$ for strong SBC,
with a continuous deformation in intermediate regimes.
This very feature shared by the equilibrium RDM considered here and the
preferred basis of decoherence is somewhat expected:
an equilibrium RDM is an asymptotic result of quantum dissipation. Due to this interesting connection, the particular diagonal representations of RDM as a result of non-canonical statistics can be also understood as one remarkable (previously overlooked) outcome
of Nature's super-selection in open quantum systems \cite{ Paz1999, Braun2001, Wang2008, Gogolin2010, Zurek2003, Zurek1981}.

\begin{figure}[h]  
	\center
	 \includegraphics[width=3.5in]{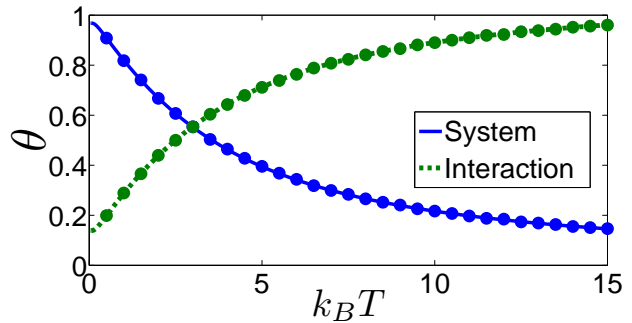}
         \caption{ Temperature dependence of the angle to be rotated on the Bloch sphere to reach eigenstates of $H_S$ (solid line) or $H_{SB}$ (dashed line) from eigenstates of equilibrium RDM, for $\gamma=0.1$, $\epsilon=0.5$ and $\omega_c=5$ (in unit of $\Delta$). Solid dots are numerically exact Monte Carlo simulations results.  }
     \label{FIG:Angle2}
\end{figure}

We now turn to the temperature dependence of non-canonical statistics at a fixed $\gamma$, as depicted in Fig.~\ref{FIG:Angle2} for $\gamma=0.1$.
We choose $\gamma=0.1$ as an example because it represents an intermediate SBC strength in Fig.~\ref{FIG:Angle1}.
As observed from Fig.~\ref{FIG:Angle2}, for temperature lower than $k_{B}T=1$ (a value considered in Fig.~\ref{FIG:Angle1}),
the RDM diagonal representation is further rotated from that of $H_S$ (solid line), but
gets closer to that of $H_{SB}$ (dashed line). Therefore non-canonical statistics becomes more pronounced
when temperature decreases. On the other hand, when temperature increases, the plotted angles continuously change in opposite directions,
showing that the RDM diagonal representation gradually moves away from the eigen-representation
of $H_{SB}$ but smoothly approaches that of $H_S$.  For temperature values
much higher than shown in Fig.~\ref{FIG:Angle2}, this trend persists.
Numerically exact Monte-Carlo simulation results (solid dots) are also
presented in Fig.~\ref{FIG:Angle2}, thus supporting again our theory.

Theoretical results outlined above may be further exploited to understand the asymptotic high-temperature behavior of RDM.
Keeping only terms of $\beta^0$ and $\beta^1$, the use of Eq.~(\ref{zerothorder}) gives  $\rho_S^{11}=1/2+\epsilon \beta/4 $
and $\rho_S^{22}=1/2-\epsilon \beta/4$; whereas the use of Eqs.~(\ref{o1}) and (\ref{o2}) yields $\rho_{S}^{12}=-\Delta \beta/4$ \cite{supple}.
But these asymptotic density matrix elements are {\it exactly} those of a canonical distribution for $H_S=\frac{\epsilon}{2}\sigma_z+\frac{\Delta}{2}\sigma_x$ at high temperature.
Non-canonical statistics is seen to have totally vanished at high temperature. This clearly depicts
an interplay between two competing factors: SBC strength and temperature.  A larger
SBC strength generates a non-canonical RDM, but the thermal averaging tends to dilute non-canonical statistics and wipes it out completely
at high temperature. Interestingly, this competition can be also appreciated from the bath statistics which is also non-canonical.
In particular,  in terms of the boson occupation number on mode $k$, the ratio of the leading-order correction to the canonical
result is proportional to $g_k^2/(\omega_k k_B T)$ \cite{supple}, which becomes negligible at high temperature.

\begin{figure}[h]  
	\center
   \includegraphics[width=3.5in]{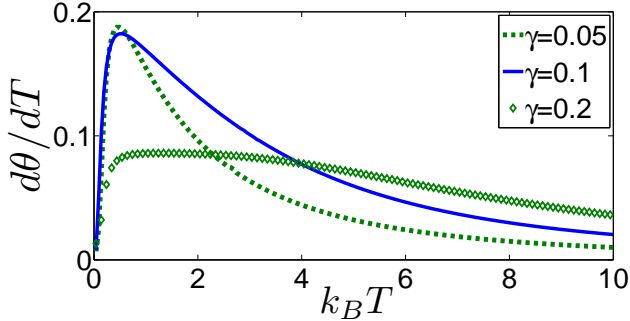}
         \caption{Sensitivity of the RDM diagonal representation to temperature variation, as described by $d\theta/dT$ vs $T$.
           $\theta$ is the angle to be rotated to reach eigenstates of $H_S$ from eigenstates of the equilibrium RDM.
           System parameters are given by  $\epsilon=0.5$ and $\omega_c=5$ (in unit of $\Delta$). }
     \label{FIG:slope}
\end{figure}

To further examine non-canonical statistics characterized by a single angle measure ($\theta$),
we present in Fig.~\ref{FIG:slope}
$d\theta/dT $, i.e.,
the sensitivity of the RDM diagonal representation to a small temperature variation,
as a function of temperature.
The sensitivity is low for very low temperature, but it rapidly increases, reaching a maximum at characteristic temperature scales that are comparable
to other system parameters (such as $\gamma$). 
As temperature increases further, the sensitivity drops to zero asymptotically.
The sensitivity profile as a function of temperature qualitatively changes for a varying SBC strength.
For weak SBC strength (e.g., $\gamma=0.05$),  it exhibits a sharp peak. Hence the rotation of the RDM diagonal representation
mainly occurs within a narrow temperature window.  For strong SBC strength (e.g., $\gamma=0.2$), the sensitivity profile displays a rather flat structure,
suggesting that it is harder for thermal effects to compete with SBC.  Thus,
if and only if the SBC strength is rather weak, then the temperature that gives $d^2\theta/dT^2=0$ (i.e., largest sensitivity of the RDM diagonal representation to
temperature) becomes an interesting temperature scale.

\section{Conclusions}
For an open quantum system not weakly coupled with a bath,
its equilibrium state is far from a canonical state at low temperature.
Exact analytical solutions are typically not available (one known exception is
the model of a harmonic oscillator linearly interacting with a boson bath).
A systematic approach to such open quantum systems
is hence highly desirable in efforts to better understand their qualitative and quantitative features of equilibrium statistics
as temperature and/or SBC strength varies.  To our knowledge, the theoretical treatment advocated in this work, as supported by numerical results,
represents the first attempt along this direction that can almost cover the whole range of SBC strength and the whole range of
temperature of a spin-boson model.
Because non-canonical statistics is closely related to strong system-bath correlation, we anticipate our theory to be also useful in understanding system-bath
entanglement.

Our theoretical findings based on a spin-boson model are much relevant to experiments based on quantum dots.
Acoustic phonon modes have been identified as the principal source of decoherence in InGaAs/GaAs quantum dots~\cite{Ramsay2010, Ramsay2010a},
and temperature is widely tunable in such a semiconductor implementation.
Certainly, in a real system the bath spectral density may not be the super-ohmic one assumed here. To address this concern we have carried out numerically
exact calculations for an Ohmic bath at a nonzero temperature,
obtaining results that are qualitatively the same as presented in this work \cite{supple} (even though our
analytical treatment based on a full polaron transformation cannot be applied to this case). Our approach can also be generalized to study any dissipative large spins as well as to ensembles of spins coupled to a common bath~\cite{General}.

\begin{acknowledgements}
 J.G. acknowledges stimulating discussions with Peter H\"{a}nggi, Guido Burkard,  Cord M\"{u}ller, and Jun-hong An.
This work is partially supported by the National Research Foundation and the Ministry of Education of Singapore.
\end{acknowledgements}

\appendix
\section{Derivations of the Off-diagonal RDM Element}
The off-diagonal element of the equilibrium reduced density matrix (RDM) can be formally written as $\rho_S^{12}=\mbox{tr}[\sigma_- \rho_t] = \mbox{tr}[\sigma_{-}D_-^2 \tilde{\rho}_t]$ where $\tilde{\rho}_t = \frac{\mbox{e}^{-\beta \tilde{H}_t}}{\mbox{tr}[\mbox{e}^{-\beta \tilde{H}_t}]} $. Expanding $\tilde{\rho}_t$ up to first order in $\tilde{H}_{SB}$, we have
\begin{eqnarray}  
		\tilde{\rho}_t &\approx& \tilde{\rho}_{t,(0)} + \tilde{\rho}_{t,(1)},
\end{eqnarray}
where
\begin{eqnarray}  
 		\tilde{\rho}_{t,(0)}  &=& \frac{\mbox{e}^{-\beta \tilde{H}_0}}{\mbox{tr}[\mbox{e}^{-\beta \tilde{H}_0}]}; \\
 		\tilde{\rho}_{t,(1)}  &=& - \frac{\mbox{e}^{-\beta \tilde{H}_0}}{\mbox{tr}[\mbox{e}^{-\beta \tilde{H}_0}]} \int^\beta_0 d\tau \mbox{e}^{\tau \tilde{H}_0} \tilde{H}_{SB} \mbox{e}^{-\tau \tilde{H}_0}, \nonumber \\
 		                      &=& - \frac{\mbox{e}^{-\beta \tilde{H}_0}}{\mbox{tr}[\mbox{e}^{-\beta \tilde{H}_0}]} \int^\beta_0 d\tau  \tilde{H}_{SB}(\tau) .
\end{eqnarray}
Inserting the above expressions into $\rho_S^{12} = \mbox{tr}[\sigma_{-}D_-^2 \tilde{\rho}_t]$, we obtain
\begin{eqnarray}  
 		\rho_{S,(0)}^{12} &=& \mbox{tr}[{\sigma_{-} D_{-}^2 \rho_{t,(0)} }] , \nonumber \\
 		                  &=& \langle \sigma_- \rangle_{\tilde{H}_S} \langle D_{-}^2 \rangle_{H_B}, \nonumber \\
 		                  &=& -\frac{B \Delta_R}{2 \eta} \tanh (\beta \eta/2);\\
 		\rho_{S,(1)}^{12} &=& \mbox{tr}[{\sigma_{-} D_{-}^2 \rho_{t,(1)} }], \nonumber\\
 		 		              &=& -\sum_{n=x,y} \int^\beta_0 d\tau \langle \tilde{H}_{SB}(\tau) \,\sigma_- D_-^2 \rangle_{\tilde{H}_0}, \nonumber \\
 		                  &=& -\sum_{n=x,y} \int^\beta_0 d\tau \langle \sigma_n(\tau) \sigma_- \rangle_{\tilde{H}_S} \langle V_n(\tau) D_-^2 \rangle_{H_B}, \nonumber \\
 		                  &=& -\sum_{n=x,y} \int^\beta_0 d\tau S_n(\tau) K_n(\tau),
\end{eqnarray}
where the explicit expressions for the system and bath correlation functions, $S_n(\tau)$ and $K_n(\tau)$, are already given in the main text.

\section{High-Temperature behavior of RDM}
Here we give some details to see how non-canonical statistics of RDM totally vanishes at high temperature.
For the off-diagonal element, the zeroth order term vanishes at high temperature since $R$ decays exponentially with $T$.
Thus, only the first-order correction term, $\rho_{S,(1)}^{12}$, contributes to the off-diagonal element at high temperature.
Furthermore, at high temperature the system correlation functions can be approximated as $S_x(\tau) \approx \frac{1}{2}\mbox{e}^{(\beta-2 \tau) \epsilon/2} $ and $S_y(\tau) \approx -i \frac{1}{2}\mbox{e}^{(\beta-2 \tau) \epsilon/2} $ (where we have used $\eta\approx \epsilon$).
Note also that though $R$ vanishes at high temperature, the term $R^2 \mbox{e}^{\phi(\tau)}$ contained in the bath correlation functions remains finite (keeping in mind that $0 \leq \tau \leq \beta$). That is,
\begin{eqnarray} 
     R^2 \mbox{e}^{\phi(\tau)}    &=&   \mbox{exp}\Big[-4 \int^\infty_0 \frac{d\omega}{\pi}              \frac {J(\omega)}{\omega^2}  \nonumber \\
     &&\ \times \frac{\cosh[\frac{1}{2}\beta  \omega]- \cosh[\frac{1}{2}(\beta -2\tau) \omega]}{\sinh(\beta  \omega/2)}\Big],\nonumber \\
              &\approx& \mbox{exp}\Big [-\kappa \frac{\tau^2- \beta \tau}{\beta} \Big],
\end{eqnarray}
where $\kappa =4 \int^\infty_0 \frac{d\omega}{\pi}\frac{J(\omega)}{\omega} =\frac{8 \gamma}{\pi} \omega_c^3$ and we have used the expansions $\sinh(x)\approx x $ and $\cosh(x) \approx 1 + \frac{1}{2} x^2$ to arrive at the second expression. The bath correlation functions at high temperature can then be written as $K_x(\tau) \approx \frac{\Delta}{4} \mbox{exp}[-\kappa \frac{\tau^2- \beta \tau}{\beta} ]$ and $K_y(\tau) \approx i\frac{\Delta}{4} \mbox{exp}[-\kappa \frac{\tau^2- \beta \tau}{\beta} ]$. We then have
$\rho_S^{12} \approx -\frac{\Delta}{4} \int^\beta_0 \mbox{e}^{(\beta-2 \tau) \epsilon/2 - \kappa(\tau^2 -\beta \tau)/\beta} d\tau $.
Since $\tau$ is also small ($0 \leq \tau \leq \beta$), the integrand can be further expanded using $\mbox{e}^{x} \approx 1+x$ and we finally obtain $\rho_S^{12} $ up to the second order in $\beta$:
\begin{eqnarray}
	 \rho_S^{12} \approx - \frac{\Delta}{4}(\beta - \kappa \beta^2 /6).
\end{eqnarray}

Calculating the diagonal elements at high temperature is more straightforward: the double integral in matrix $A$ [see Eq.~(12)] indicates that it is at least proportional to $\beta^2$ and can be discarded if we are only interested in terms up to the first-order of $\beta$.
The diagonal elements, $\rho_S^{11}$ and $\rho_S^{22}$, can then be written as
$\frac{1}{2}(1\pm \mbox{tr}_S[\sigma_z \tilde{\rho}_S^{(0)}])$. The are explicitly given by
\begin{eqnarray}
	 \rho_S^{11} &=& \frac{1}{2}[1-\frac{\epsilon}{ \eta} \tanh(\beta \eta /2)] \approx \frac{1}{2}(1- \epsilon \beta/2); \\
	 \rho_S^{22} &=& \frac{1}{2}[1+\frac{\epsilon}{ \eta} \tanh(\beta \eta /2)] \approx \frac{1}{2}(1+ \epsilon \beta/2),
\end{eqnarray}
where we have used $\eta=\sqrt{\epsilon^2+\Delta_R^2} \approx \epsilon $ and $\tanh(x) \approx x$.

Gathering all the results above, our analytic theory predicts that at high temperature,
\begin{eqnarray}
         \rho_{S} &=&  \frac{1}{2}\begin{pmatrix} 1 -  \frac{\epsilon \beta }{2} &  -\frac{\Delta \beta }{2} \\ -\frac{\Delta \beta }{2} & 1 +  \frac{\epsilon \beta }{2} \end{pmatrix}.
\end{eqnarray}
The above expression is exactly the same as the high-temperature canonical state of the system, $\frac{\mbox{e}^{-\beta H_S}}{\mbox{tr}_S[\mbox{e}^{-\beta H_S}]}$. This remarkable agreement nicely demonstrates that canonical statistics is recovered at high temperature.

\section{Ohmic Bath}

\begin{figure}[h]  
	\center
   \includegraphics[width=3.5in]{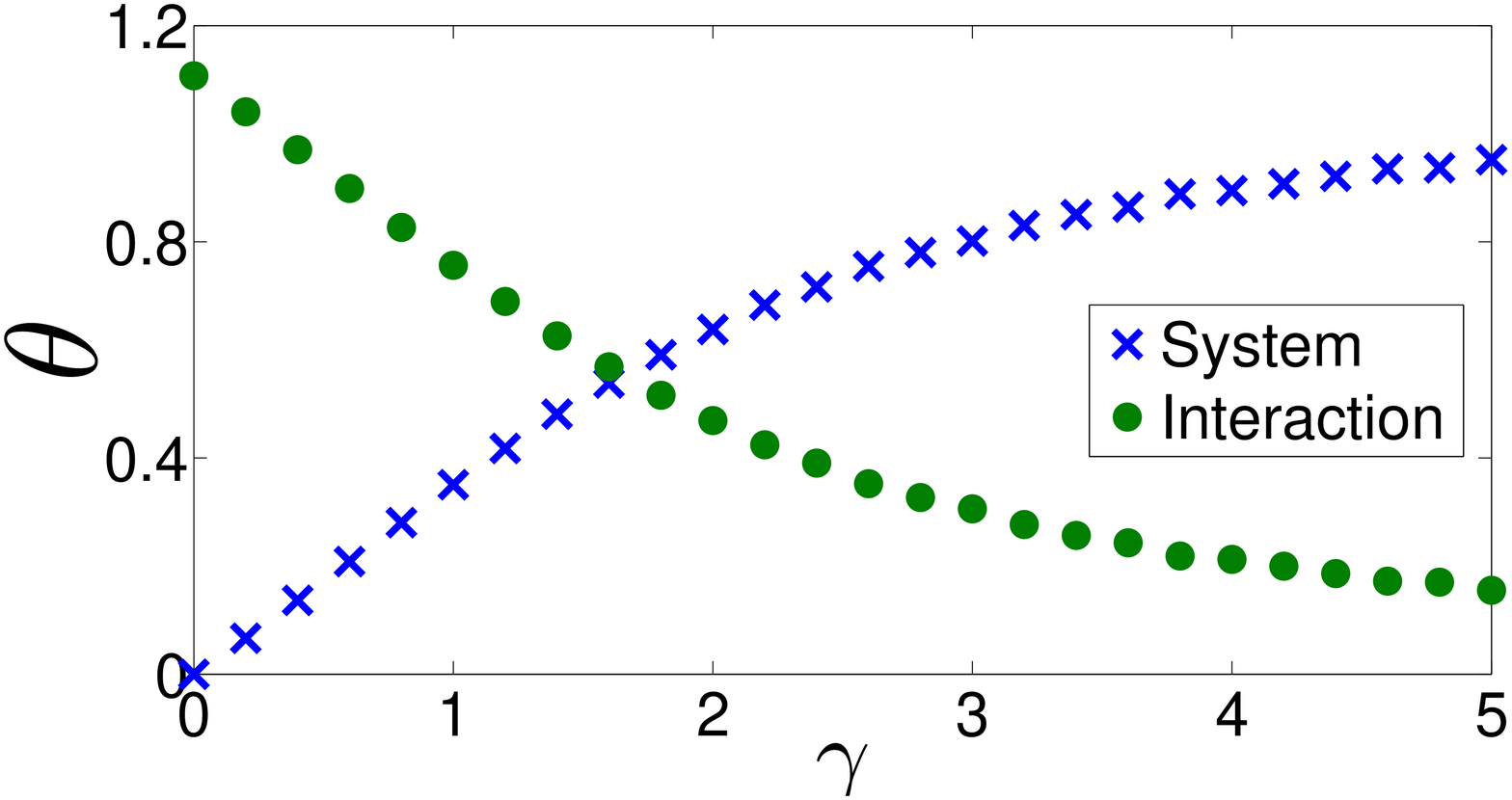}
         \caption{ The angle to be rotated on the Bloch sphere to reach the eigenstates of $H_S$(crosses) or $H_{SB}$ (solid dots) from the eigenstates of the equilibrium RDM as a function of coupling strength, for $\beta=1$, $\epsilon=0.5$ and $\omega_c=5$ (in unit of $\Delta$). }
     \label{FIG:ohmic1}
\end{figure}

\begin{figure}[h]  
	\center
   \includegraphics[width=3.5in]{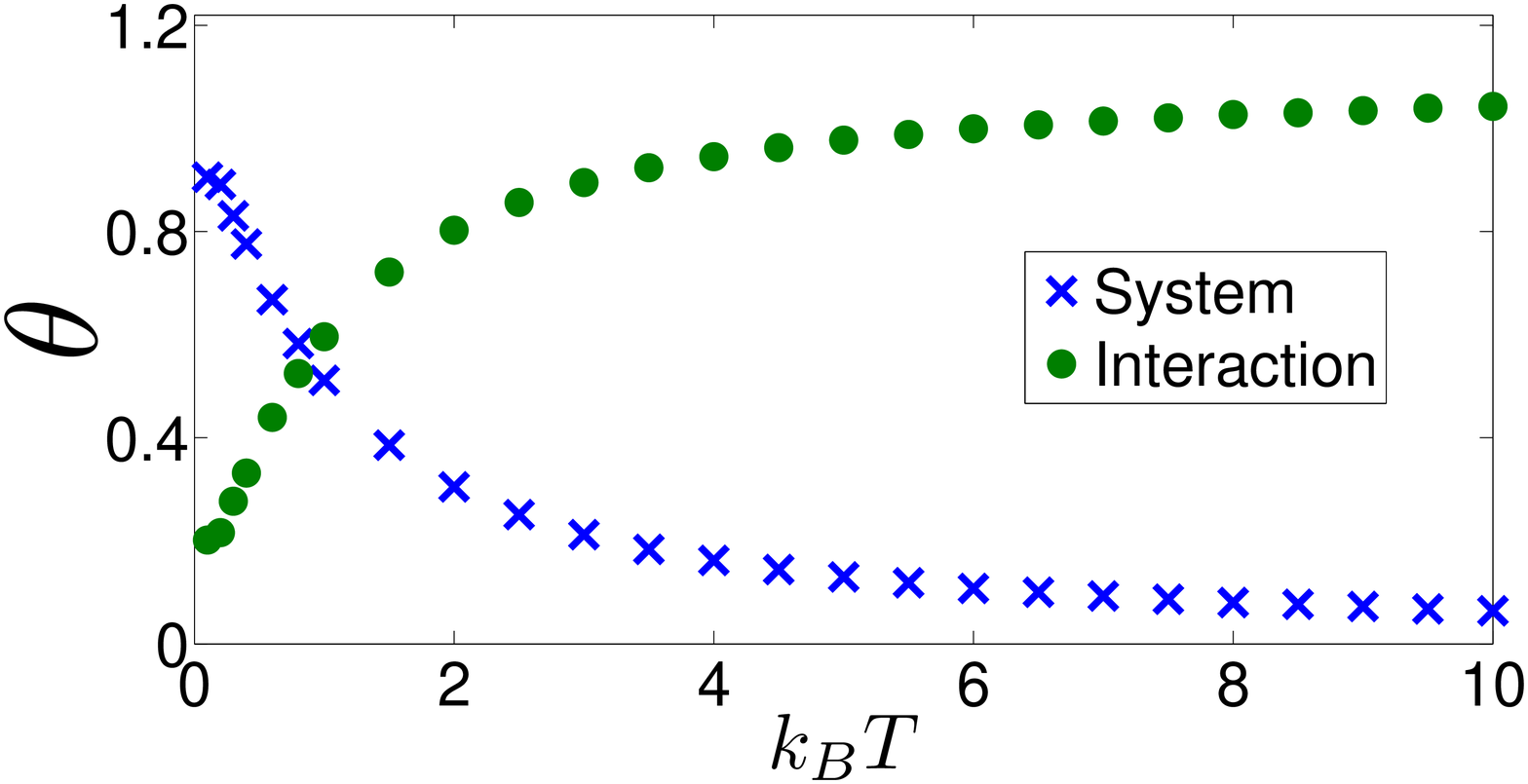}
         \caption{  The angle to be rotated in the Bloch sphere to reach the eigenstates of $H_S$(crosses) or $H_{SB}$ (solid dots) from the eigenstates of the equilibrium RDM as a function of temperature, for $\gamma=1.5$, $\epsilon=0.5$ and $\omega_c=5$ (in unit of $\Delta$). }
     \label{FIG:ohmic2}
\end{figure}

Here we study the rotation angle between the eigenstates of the RDM and the eigenstates of $H_S$ or $H_{SB}$ using an Ohmic bath, $J(\omega)=\gamma \omega \mbox{e}^{-\omega/\omega_c}$. Unfortunately, a full polaron method as we used in the main text
is not applicable for an ohmic bath as it suffers from an unphysical divergence issue~\cite{Leggett1987, weiss2008}. The integral in the renormalization constant $R$ is divergent for all coupling strength, and the tunneling element is always normalized to zero. Therefore here we only present the numerical results from the imaginary time path integral simulations (for not too low temperature). The coupling and temperature dependence of the rotation angle are plotted in Fig.~\ref{FIG:ohmic1} and Fig.~\ref{FIG:ohmic2}. It can be seen that the features of the figures are qualitatively similar to those obtained using a super-ohmic spectral density in the main text. Therefore, the general observations made in our main
text should not be sensitive to the spectral density of the bath.

\section{Eigenvalue at Zero Temperature}
In the eigenbasis, the equilibrium RDM can be written as
$\rho_S = \left( \begin{smallmatrix} \lambda_1& 0 \\ 0& \lambda_2 \end{smallmatrix}\right)$,
where $\lambda_i$ are the eigenvalues of RDM and $\lambda_1 + \lambda_2=1$. The eigenvalues denote the population of each of the eigenstate. The eigenvalues also serve as an indicator of the purity of the system. If both eigenvalues are non-zero, the system is in a mixed state. Below, we will use the larger eigenvalue, $\lambda_2$, to investigate the purity of the system: the system is in a pure state if $\lambda_2=1$ and vice versa.

\begin{figure}[h!]  
	\center
   \includegraphics[width=3.0in]{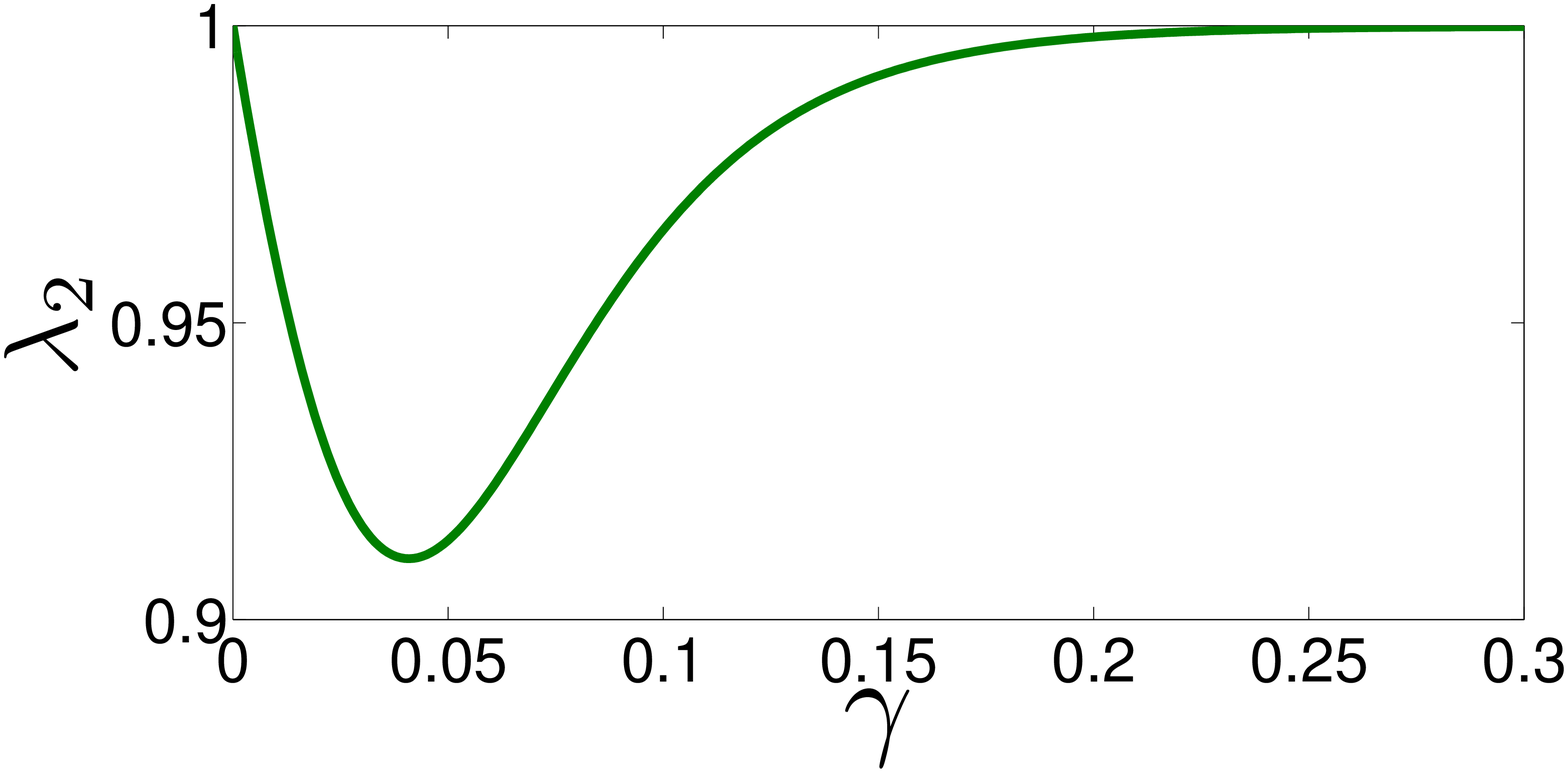}
         \caption{ The larger eigenvalue, $\lambda_2$ of RDM, plotted against the SBC strength $\gamma$, for $T=0$, $\epsilon=0.5$ and $\omega_c=5$ (in unit of $\Delta$). }
     \label{FIG:eigenvalue}
\end{figure}

Due to the finite system-bath coupling, the equilibrium RDM might not be a pure state at $T=0$ even though the system plus the bath is in their entangled ground state. To examine how the purity of RDM at $T=0$ depends on the coupling strength, we plot $\lambda_2$ as a function of $\gamma$ in Fig.~\ref{FIG:eigenvalue}.
It can be observed that the eigenvalue exhibits an interesting non-monotonic behavior as a function of the system-environment
coupling strength. At $\gamma=0$, the system is in the ground state of $H_S$ with unit purity. At finite coupling, both eigenstates are populated and RDM is a statistical mixture due to the system-bath entanglement. Interestingly, RDM is reduced to a pure state at very large $\gamma$, indicating that the system-bath entanglement vanishes at ultra-strong system-environment coupling. However this pure state is no longer the eigenstate of the system Hamiltonian, but that of $\sigma_z$ in the interaction Hamiltonian.

\section{Bath Statistics}
Here we examine the equilibrium statistics of the bath by examining the average boson number of each mode, which is denoted by
 $\langle n_k \rangle$. In the polaron frame, the boson number operator is given by
\begin{eqnarray}
\tilde{n}_k &=& \mbox{e}^{F} n_k \mbox{e}^{-F}, \nonumber\\
		    &=& n_k - \frac{g_k}{\omega_k}\sigma_z (b_k^\dagger + b_k) + \frac{g_k^2}{\omega_k^2}, \label{eqn:n_polaron}
\end{eqnarray}
where $F=\sigma_z \sum_k\frac{g_k}{\omega_k}(b_k^\dagger - b_k)$ in the first line. An approximate expression of $\langle n_k \rangle$ can be obtained by
\begin{eqnarray}
		\langle n_k \rangle &=& \frac{\mbox{tr} [n_k\,\, \mbox{e}^{-\beta H_t}]}{ \mbox{tr} [\mbox{e}^{-\beta H_t]}}, \nonumber\\
		                    &=& \frac{\mbox{tr} [ \mbox{e}^{F}  n_k \mbox{e}^{-F} \,\,\mbox{e}^{F} \mbox{e}^{-\beta H_t} \mbox{e}^{-F}]}{\mbox{tr} [\mbox{e}^{-\beta H_t}]},\nonumber \\
		                    &=& \frac{\mbox{tr} [\tilde{n}_k\mbox{e}^{-\beta \tilde{H}_t}]}{\mbox{tr} [\mbox{e}^{-\beta H_t}]},\nonumber \\
		                    &\approx&  \frac{\mbox{tr} [\tilde{n}_k\mbox{e}^{-\beta \tilde{H}_0}]}{\mbox{tr} [\mbox{e}^{-\beta H_0}]}.
\end{eqnarray}
Inserting Eqn.~(\ref{eqn:n_polaron}) into the above expression, we have
\begin{eqnarray}
	\langle n_k \rangle= \langle n_k\rangle_{{0}} + g_k^2/\omega_k^2,
\end{eqnarray}
where $\langle n_k\rangle_{{0}}= (\mbox{e}^{\frac{\omega_k}{k_B T}}-1)^{-1}$ is the average boson number without system-bath coupling. In the high temperature limit, it satisfies the equipartition theorem, $\langle n_k\rangle_{{0}} \approx \frac{k_B T}{\omega_k}$. Therefore, the fractional correction,
given by $\frac{\langle n_k \rangle- \langle n_k \rangle_{0}}{\langle n_k \rangle_{0}}\approx \frac{g_k^2}{\omega_k k_B T}$, becomes negligible at high temperature.


\begin{thebibliography}
\bibitem{correlation-function-note} We use $\langle \cdot\rangle_{{H}_B}$ to denote an average over
$\frac{\mbox{e}^{-\beta {H}_B}}{ \mbox{tr}_B [\mbox{e}^{-\beta {H}_B}]}$. Similarly, later
we use $\langle \cdot\rangle_{\tilde{H}_S}$ denoting an average over
        $\frac{\mbox{e}^{-\beta \tilde{H}_S}}{ \mbox{tr}_S [\mbox{e}^{-\beta \tilde{H}_S}]}$.
\bibitem{phi-note3} Using a super-ohmic spectral density, we have $\phi(\tau) = \frac{4\gamma}{\pi \beta^2} \Big[ \psi'\Big(\frac{1+(\beta-\tau)\omega_c}{\beta \omega_c} \Big) + \psi'\Big(\frac{1+\tau\omega_c}{\beta \omega_c} \Big)\Big]$.
\bibitem{supple} See Appendix for necessary details
of our theoretical derivations, high-temperature behavior of RDM,
additional numerical results for a different bath spectrum density, dicussions on the eigenvalues of RDM as a function of system parameters,
as well as the bath statistics.
\bibitem{footnote} For a spin-boson model with sub-ohmic or ohmic spectral density, one may expect an abrupt phase transition at very low
temperature.
\end{thebibliography}

\end{document}